# Estimating saturated hydraulic conductivity from surface Ground-penetrating radar monitoring of infiltration

Emmanuel Léger[1], Albane Saintenoy[2] and Yves Coquet[3]

[1] Université Paris Sud, UMR8148 IDES, Orsay, France, emmanuel.leger@u-psud.fr
[2] Université Paris Sud, UMR8148 IDES, Orsay, France, albane.saintenoy@u-psud.fr
[3] Université Orléans, ISTO/OSUC.,Orléans, France, yves.coquet@univ-orleans.fr

**Abstract**

In this study we used Hydrus-1D to simulate water infiltration inside a ring infiltrometer. We generated water content profiles at each time step of infiltration, based on a particular value of saturated hydraulic conductivity, knowing the other van Genuchten parameters. The water content profiles were converted to dielectric permittivity profiles using the Complex Refractive Index Method relation. We then used the GprMax suite of programs to generate radargrams and to follow the wetting front using arrival time of electromagnetic waves recorded by a Ground-Penetrating Radar (GPR). Theoretically the depth of the inflection point of the water content profile simulated at any infiltration time step, is related to the peak of the reflected amplitude recorded in the corresponding trace in the radargram. We used this relationship to invert the saturated hydraulic conductivity for constant and falling head infiltrations. We present our method on synthetic examples and on two experimental results acquired on sand. We further discuss on the possibility of estimating two other van Genuchten parameters, $n$ and $α$, in addition to the saturated hydraulic conductivity.

1. Introduction

Soil hydraulic properties described by the soil water retention $θ(h)$ and hydraulic conductivity $K(h)$ functions, dictate water flow in the vadose zone, as well as the partitioning between infiltration and runoff. Their evaluation has important applications for modelling available water resources and for flood forecasting. It is also crucial in evaluating soil capacity to retain chemical pollutants and in assessing the potential of groundwater pollution.

The determination of the parameters involved in the van Genuchten soil water retention function (van Genuchten, 1980) is usually done by laboratory experiments, such as the water hanging column (Dane and Hopmans, 2002). Hydraulic conductivity, on the other hand can be estimated either in laboratory, or *in situ* using infiltrometry tests. Among the large panel of existing tests (Angulo-Jaramillo, 2002), the single (Müntz *et al.*, 1905) or double ring infiltrometers (Boivin *et al.*, 1987) give the *field* saturated hydraulic conductivity by applying a positive charge on soils, whereas the disk infiltrometer (Perroux et White, 1988; Clothier and White, 1981) allows to reconstruct the hydraulic conductivity curve, by applying different charges smaller than or equal to zero.

In infiltrometry tests, the volume of infiltrated water versus time is fitted to infer soil hydraulic conductivity at or close to saturation. These tests are time-consuming and difficult to apply to landscape-scale forecasting of infiltration. Furthermore they involve various simplifying assumptions, partly due to the ignorance of the infiltration bulb shape.

Starting from this issue, geophysical methods have been applied to the infiltration process, mainly electrical (Battle-Aguilar et al., 2009) and electromagnetic methods. Among them, Ground-Penetrating Radar (GPR) is based on electromagnetic (EM) wave propagation. It is highly sensitive to water content variations directly related to the dielectric permittivity (Huismann et al., 2003). Thus it appears to be an accurate tool for wetting front infiltration monitoring (Saintenoy et al., 2008).

## 2. Method

We studied the infiltration of a 5-cm thick water layer inside a single ring infiltrometer, in a sandy soil. The draft of the apparatus is presented in Fig. 1. The single ring infiltrometer is a 1-mm thick aluminum cylinder of 60-cm diameter, approximately 20-cm high, buried in the soil by 10 cm. We set GPR antennae (namely the transmitter T and the receiver R) at a variable distance of the edge of the cylinder, noted x in Fig. 1. In all our field experiment, we used a Mala RAMAC system with antennae centered on 1600 MHz, shielded at the top. We then covered the inner part of the cylinder with a plastic waterproof sheet. The plastic sheet allowed us to fill in the cylinder with water, and create an initial 5-cm thick layer of water, preventing infiltration into the sand, before starting the acquisition. The beginning of the acquisition was launched by pulling away the plastic sheet to trigger the water infiltration. The GPR system was set to acquire a trace every 10 s. With this apparatus, we performed two types of infiltration: i) a falling head infiltration consisting in pulling away the plastic sheet and leaving the water to infiltrate the sand freely with no refill, ii) a constant head case, in which water was added inside the ring to maintain a 5-cm thick layer of water constantly during the infiltration experiment. In the following examples we will show how we can use the GPR data acquired every 10 s during the infiltration to get the hydraulic conductivity at saturation.

## 3. Falling head infiltration case

### 3.1. Numerical example

*Forward modelling*

The falling head infiltration case was simulated using Hydrus-1D (Šimůnek *et al.,* 1996, 2008; Šimůnek and van Genuchten, 1996) with a model made of a homogeneous medium of 60 cm diameter and 50 cm depth, divided into 1001 layers. To describe the soil hydraulic properties of the medium, we used the van Genuchten-Mualem (van Genuchten, 1980) hydraulic conductivity and water retention functions, which gives us 5 parameters, namely the water content at saturation $\theta_s$, the residual water content $\theta_r$, two fitting parameters $\alpha$ and $n$, and the hydraulic conductivity at saturation $K_s$. For our numerical example we set $\theta_s=0.43$, $\theta_r=0.07$, $\alpha=0.019$ cm$^{-1}$, $n=8.67$ and $K_s=0.120$ cm/min. We used an atmospheric boundary condition with no rain and no evaporation and a free drainage for the bottom. To simulate the 5-cm layer of water, the water potential initial condition was set to 5 cm for the top node. We simulated the first 10 minutes of the experiment with a time step of 10 s, i.e. with 60 water content snapshots. Using the CRIM relation (Birchak *et al.*, 1974; Roth *et al.*, 1990), each water content snapshot was converted to permittivity profiles (made of 1001 points), considering a three-phase media: sand (considered as pure silica), water and air. Each one of these permittivity profiles (Fig. 2-a) were the input of the GprMax2D program (Giannopoulos, 2004), based on finite difference time domain modeling, which output is one simulated GPR trace acquired at the antenna position (set at the surface of the medium, in the middle with X=0.3 m in Fig. 1). The simulated GPR monitoring of the infiltration is shown in Fig. 2-b. The horizontal axis is the number of traces simulated by GprMax2D, two traces being separated by 10 seconds, as Hydrus-1D profiles are. The vertical axis is the Two-Way Traveltime (TWT) of the EM wave amplitude coming back to the receiver.

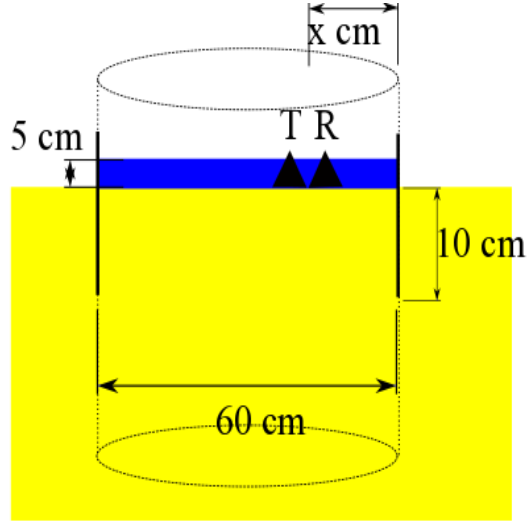

Figure 1. Drafts of the apparatus at initial state.

On the profile presented in Fig. 2-b, we denote one particular reflection, labelled A. Its arrival time is increasing as the wetting front goes deeper. The reflection is interpreted as coming from the wetting front. The reflections labelled A' and A" are primary and secondary multiple from reflection A. The reflection labelled B is the wave traveling in air directly between the two antennae. After the 40$^{th}$ trace, the 5-cm layer of water has been infiltrated, and drainage is starting. As a consequence, the permittivity of the upper part of the medium decreases. The EM wave velocity is inversely proportionnal to the square root of permittivity (Huisman *et al.*, 2003). Then the TWT of reflection A increases more slowly, creating a change of slope in the reflection time curve (Fig. 2-b). On Fig. 2-c, we represent two curves: the TWT of the maximum peak of reflection A picked on Fig.2-b and the TWT calculated by ray-path algorithm, going from the GPR antennae to the inflection point of $\varepsilon(z)$ curves (crosses on Fig. 2-a) as proposed by Saintenoy and Hopmans ( 2011). We attribute the difference between those two curves to some numerical dispersion of the signal, problem that we will address later.

*Inversion*

We used the TWT picked from the radargram of Fig. 2-b as data to be fitted in the idea of going back to the hydraulic conductivity at saturation, assuming the other 4 parameters are known. Using Hydrus1D, we generated the 60 water content snapshots using a $K_s$ value in the range 0.01 to 1 cm/min with a step of 0.001 cm/min. For each value of $K_s$, we calculated the TWT by the ray-path algorithm from the surface to the inflection point of the $\varepsilon(z)$ curves. We computed the Root Mean Square Error (RMSE) between these times and the data. It was minimized for $K_s = 0.129$ cm/min. This value is higher than the one taken for simulating the data, i.e. $K_s = 0.12$ cm/min.

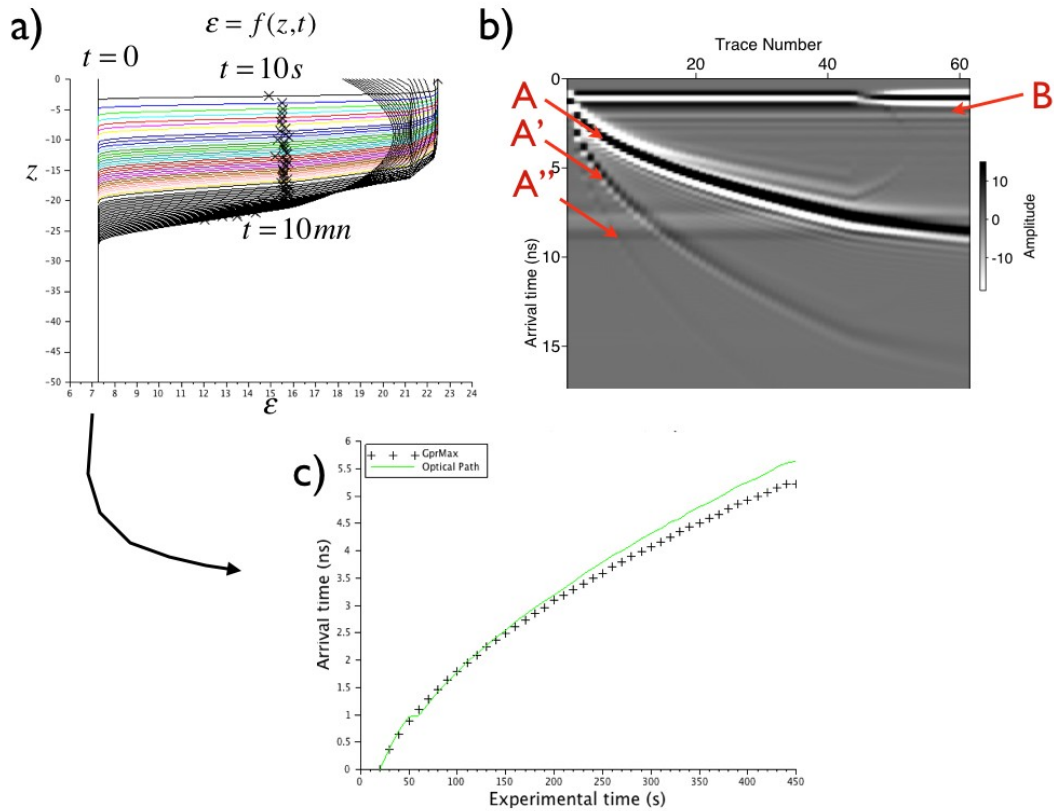

Figure 2. Falling head infiltration of a 5-cm thick water layer. a) Permittivity profiles: each curves is plotted every 10 s; Crosses represent the inflection points of each curve. b) Radargram simulated with GprMax2D, reflection A is coming from the wetting front, B is the direct wave, A' and A" are multiples of reflection A. c) TWT computed with optical path algorithm, directly from the permittivity profiles.

## 4. Experimental example

The experiment took place in a quarry of Fontainebleau sand in Cernay-La-Ville (Yvelines, France). The middle of the antennae were positioned at 11 cm away from the cylinder (x = 11 cm in Fig. 1). The 5-cm water layer was totally infiltrated after 10 minutes but in certain areas of the infiltration surface this time has been slightly shorter. The recorded GPR data are shown in Fig. 3. We recorded during 30 minutes, with a time window of 15 ns, transmitting and receiving each 10 seconds (stacking 4 measurements). We subtracted the average trace and applied an Automatic Gain Control to the data. The sand parameters have been determined in laboratory by several classical hanging water column experiments, fitted by the van Genuchten retention curve. Assuming a 5% uncertainty on the optimized parameters, we obtained $\theta_r = 0.062 \pm 0.003$, $\theta_s = 0.39 \pm 0.01$, $\alpha = 0.023 \pm 0.001$ 1/cm and $n = 6.7 \pm 0.3$. We considered in our model that the sand was homogeneous. Gravimetric measurements on field samples gave us an initial volumic water content of $\theta_i = 0.09 \pm 0.01$. On the profile presented in Fig. 3, we denoted three particular reflections. The one interpreted as coming from the infiltration front, labelled A, is visible on the first 30 minutes of the acquisition with an arrival time varying from 2 ns down to 9 ns. The other reflections are interpreted in Léger and Saintenoy (2012).

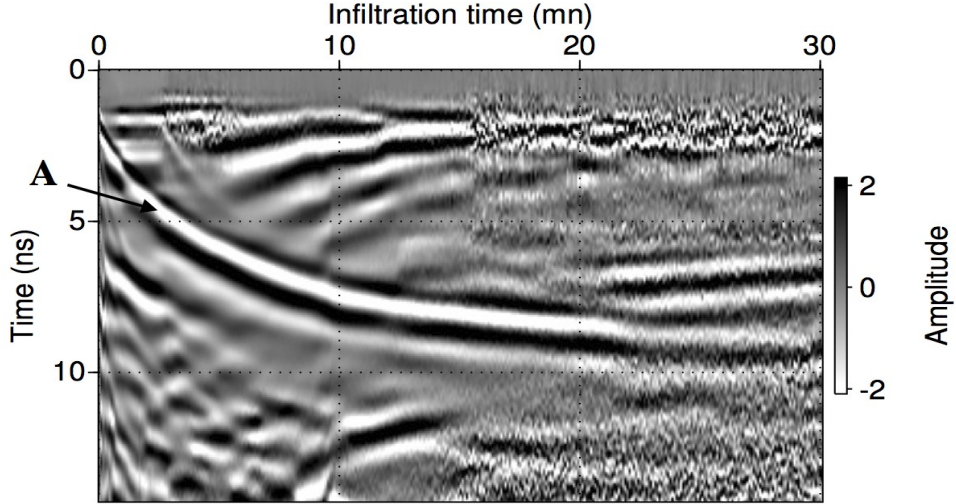

Figure 3. Experimental GPR data acquired during the unconstant head infiltration of 5 cm. Reflection A is the reflection coming from the wetting front.

We picked the arrival time of the A reflection peak and inverted the saturated hydraulic conductivity based on the same algorithm as presented in the synthetic case. We obtained the minimum of the objective function for $K_s = 0.131$ cm/min. In parallel we have made disk infiltrometer experiments, using the multi-potential method (Ankeny et al, 1991; Reynolds and Elrick, 1991). We obtained a value of saturated hydraulic conductivity of $K_{DISK} = 0.108 \pm 0.01$ cm/min.

*Uncertainty analysis*

We assumed a 5% relative uncertainty on the four van Genuchten parameters and the initial water content of the sand. These parameters influence directly the arrival time curves used to compute the misfit function. Fig. 4 shows variations associated with each of those parameters. It appears that the uncertainty on $\alpha$ has the strongest influence on the arrival time curves. As an attempt to evaluate the uncertainty in the hydraulic conductivity at saturation retrieved from GPR data fitting, we made some quadratic error summation associated with each of the parameters. The total quadratic error has the expression of $\delta_{tot} = \sqrt{\delta_{ti}^2 + \delta_{tr}^2 + \delta_{ts}^2 + \delta_n^2 + \delta_\alpha^2 + \delta_{GPR}^2 + \delta_{algo}^2}$, where $\delta_{ti}, \delta_{tr}, \delta_s, \delta_n, \delta_\alpha$ are RMSE due to uncertainties on $\theta_i, \theta_r, \theta_s, n, \alpha$ presented above, those RMSE being the summation of the two curves for each parameters, presented on Fig. 4.

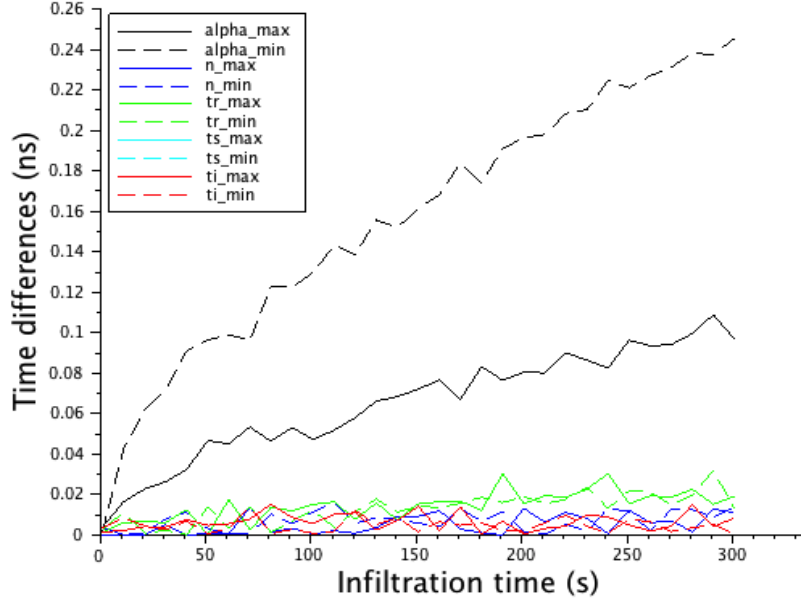

Figure 4. Influence of the parameter uncertainties on the arrival time difference, between the arrival time generated with each paramaters and the time generated with optimal parameters.

$\delta_{GPR}=0.1$ ns is assumed to be the error on picking, and $\delta_{algo}=0.010$ ns is the RMSE between the arrival times generated by the optical path-algorithm and by the GprMax modelling (Fig. 2-c).

The total quadratic error was estimated at $\delta_{tot}=0.111$ ns. From the objective function curve, all the $K_s$ in the interval $[0.04\,;0.263]$ cm/min has a misfit value to our data less than 0,111 ns, with the minimum at 0.131 cm/min. We think that this wide interval of possible $K_s$ is over-estimated by our rough error determination.

## 5. Constant head infiltration case

### 5.1. Numerical example

In this second case, the infiltration takes place in the field keeping the level of water constant, 5 cm above the ground during the whole experiment. By the same way as presented above, using the same van Genuchten parameters as in the first synthetic example ( $\theta_s=0.43$, $\theta_r=0.07$, $\alpha=0.019$ cm$^{-1}$, $n=8.67$ and $K_s=0.120$ cm/min), we modeled this infiltration of water inside a ring infiltrometer by applying a constant head of 5 cm for top node boundary condition during 10 minutes. The permittivity profiles are presented on Fig. 5-a, each curve is plotted every 10 s as the previous case. The Fig. 5-b shows the radargram simulated with GprMax 2D. As it can be seen, the reflexion labelled A, coming from the infiltration front is always coming later and later, because the infiltration is constantly fed by the constant ponding depth, contrary to the previous unconstant head case. On Fig. 5-c), we computed the TWT of the wetting front, using ray path algorithm and the picking of A reflexion coming from the radargram Fig. 5-b.

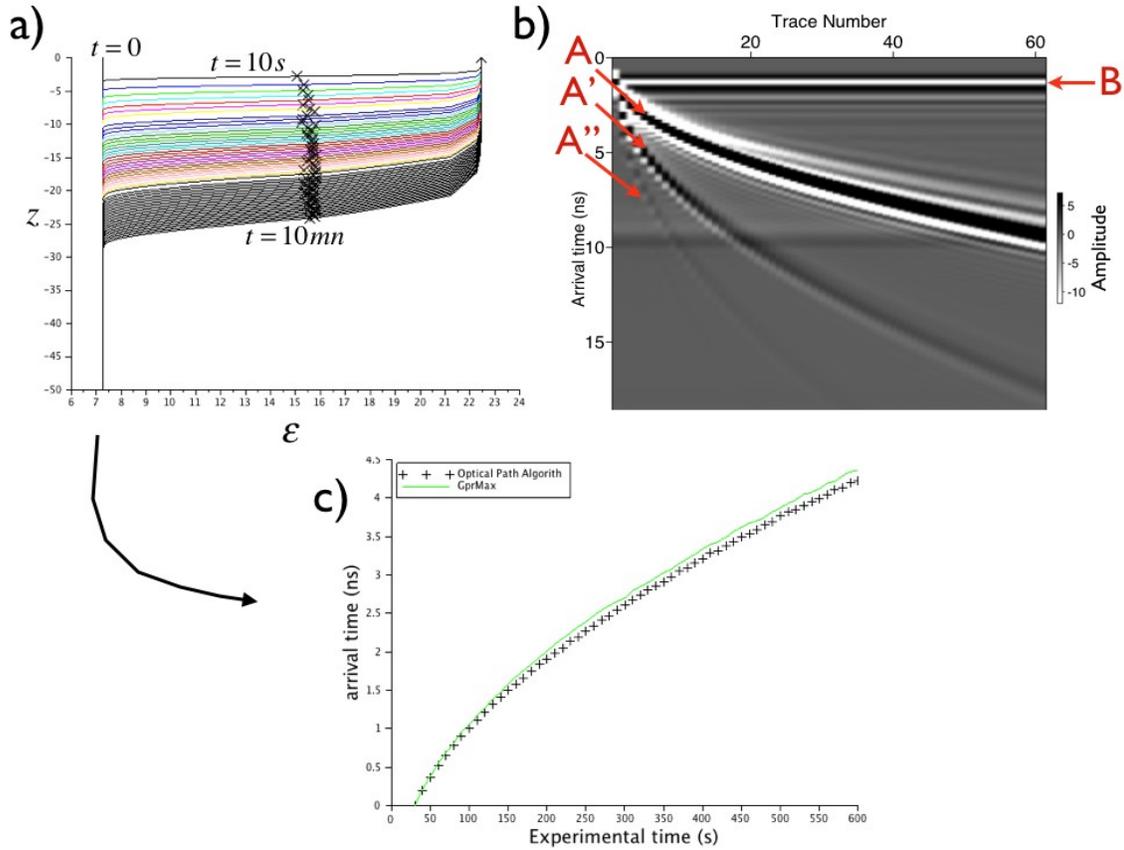

Figure 5. Constant head infiltration of 5 cm of water. -a) Permittivity profiles, each curves is plotted every 10 s. Crosses represent the inflection points. -b) Radargram simulated with GprMax-2D, reflection A is the wetting front, B is the direct wave, A' and A" are shadows. -c) Two Way travel Time computed with optical path algorithm, directly from the permittivity profiles.

Assuming the four van Genuchten parameters $\theta_s, \theta_r, \alpha, n$ known, we inverted for the saturated hydraulic conductivity, by minimizing the differences between the arrival times of the wetting front reflection obtained by ray path-algorithm and the arrival times picked from the radargram Fig. 5-b. The objective function was minimized for $K_s = 0.128$ cm/min, this value being again higher than the one for simulating the data: $K_s = 0.120$ cm/min.

### 5.2. *Experimental example*

The experiment took place in the same quarry of Fontainebleau sand as the previous experiment. The middle of antennae was positioned in the middle of the ring (x = 30 cm in Fig.1). The recorded GPR data are shown on Fig. 6. We recorded during 80 minutes, with a time window of 30 ns (in Fig. 6, we only present a part of the radargram), transmitting and receiving each 5 seconds, with no stacking. In the radargram of Fig. 6, we subtracted the average trace and apply an AGC gain to the data. We used the van Genuchten parameters coming from the laboratory hanging column experiments and we measured on sand samples an initial volumetric water content of $\theta_i = 0.07 \pm 0.02$.

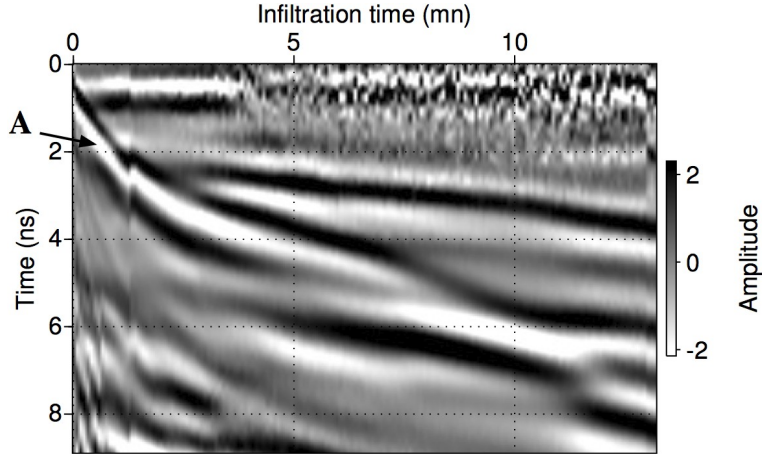

Figure 6. GPR data acquired during a constant head (5 cm) infiltration.
Reflection A is the reflection coming from the wetting front.

On the profile presented in Fig. 6, the arrival time of reflection A ranges from 0 at the beginning of the experiment to about 6 ns after 10 mn. We picked the arrival time of the reflection A peak and computed the objective function using the same procedure described before. We obtained the minimum of the objective function for $K_s = 0.110$ cm/min. This value has to be again compared with the one obtained by the disk infiltrometer experiment, $K_{DISK} = 0.108 \pm 0.01$ cm/min.

By the same procedure presented in the former field example, we found a total quadratic error of $\delta_{tot} = 0.131$ ns, which gives a range of possible values for the saturated hydraulic conductivity, $K_s = [0.035 ; 0.213]$ cm/min.

## 6. Toward a three parameter inversion

From our uncertainty analysis (Fig. 4), we pointed out the high sensitivity of our inversion to the parameter $\alpha$. The solution might be in inverting from our GPR measurements for $\alpha$ and $n$ in addition to $K_s$, assuming $\theta_r$ and $\theta_s$ known. We computed misfit diagrams of the arrival times, to check for which parameters would be easily invertible or not by our algorithm. The principle of our misfit diagrams is presented on Fig. 7-d. We carried on this analysis in both infiltration cases. We set one of the three parameters and plotted the misfit diagram for the two others. In Fig. 7, we present the obtained misfits diagrams for the constant head infiltration. We obtained similar ones for the falling head case. On the graphics of Fig. 7, for each time Hydrus-1D cannot resolve the Richards equation for a set of parameters, we return the same misfit equal to 10 (dark red areas on Fig. 7-a to c). The misfit diagram $(K_s, \alpha)$ on Fig. 7-b exhibits a correlation between those two parameters, and informs us on the possibility of inverting $(K_s, \alpha)$ knowing $n$. The two other diagrams show the difficulty of obtaining the correct value for n. Indeed this parameter is directly related to the slope of the retention curve $\frac{\partial \theta}{\partial h}$ thus as well as to $\frac{\partial \theta}{\partial z}$ at the infiltration, which information is not present in the TWT of the inflection points, used for the misfit computation.

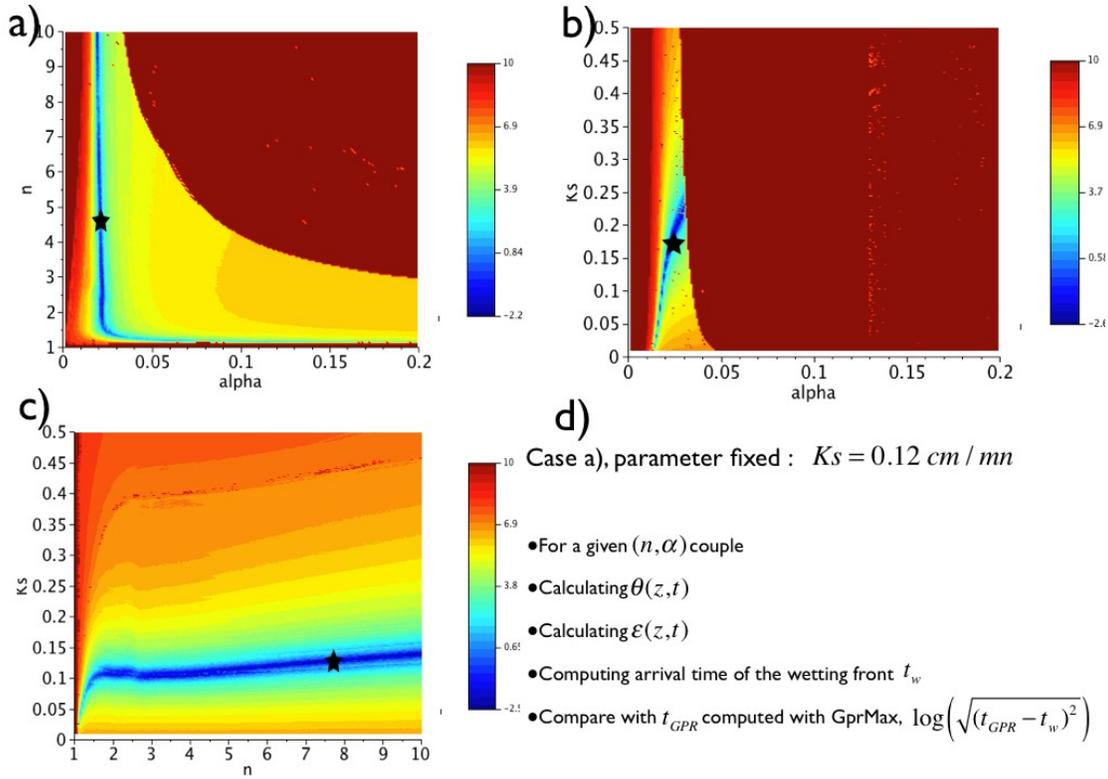

Figure 7: Misfit diagram for the case constant head boundary condition for top Nodes. a) $(n, \alpha)$ diagram, b) $(K_s, \alpha)$ diagram, c) $(K_s, n)$ diagram. Blacks stars represent minimum.

As a last case, we present an other type of falling head infiltration, but in this case, variation of the level of water is coming from an experiment. The Hydrus-1D top node condition was set to variable boundary condition. We present the misfit diagrams in Fig 8, the synthetical inversion being the same as other cases. Note again on Fig 8-a to 8-c that each time that Hydrus-1D does not converge, we return the same misfit (dark red zones on the graphics). Graphics 8-a) and 8-c) are completely different from the previous cases. They imply that adding information on the water layer thickness evolution from the experiment itself is bringing some information that could help to resolve for the n parameter. This case needs to be explored in more details.

**References**

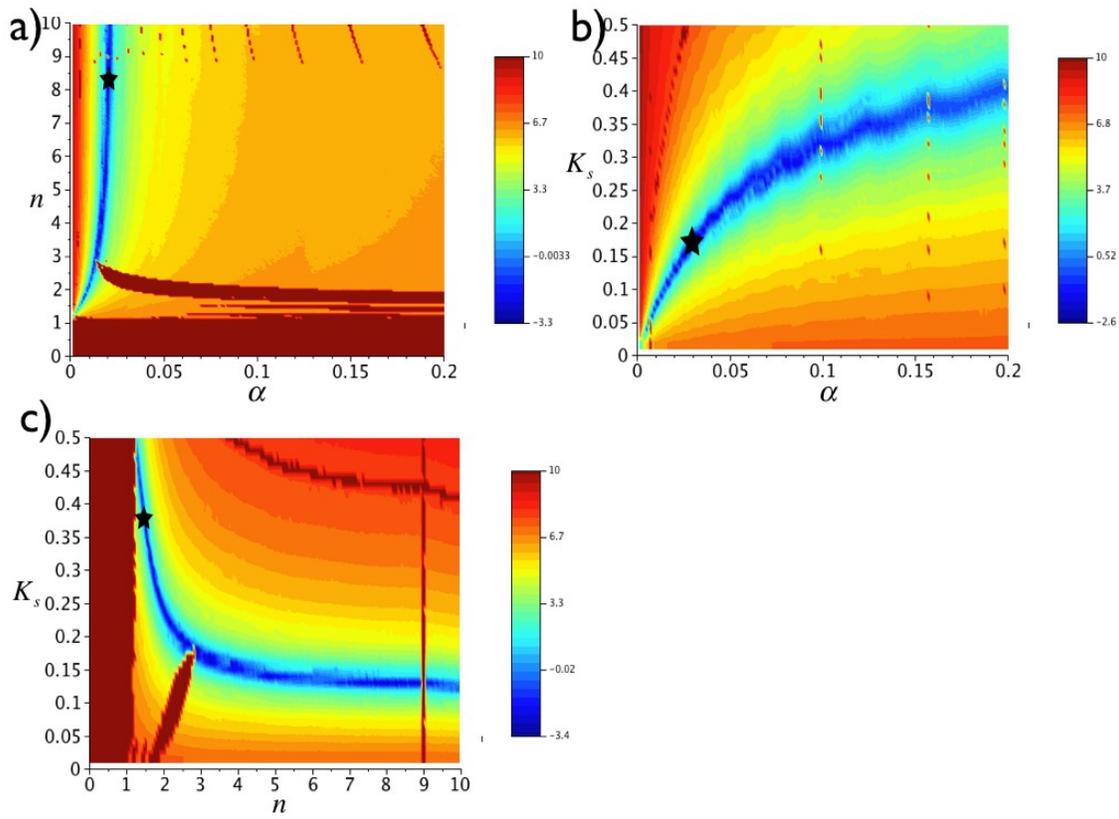

Figure 8: Misfit diagram for the case Unconstant head boundary condition, set by experiment, for top node. a) $(n, \alpha)$ diagram, b) $(K_s, \alpha)$ diagram, c) $(K_s, n)$ diagram. Blacks stars represent minimum.

Angulo-Jaramillo, R., J. Vandervaere, S. Roulier, J. Thony, J. Gaudet, and M. Vauclin, Field measurement of soil surface hydraulic properties by disc and ring infiltrometers. a review and recent developments, *Soil and Tillage Research*, no. 55, pp. 1–29, 2000.

Ankeny, M.D., M. Ahmed, T.C. Kaspar, and R. Horton, Simple field method for determining unsaturated hydraulic conductivity. Soil Sci. Sco. Am. J., 55, 467-470, 1991.

Battle-Aguilar, J., Schneider, S., Pessel, M., Tucholka, P., Coquet, Y. and P. Vachier, Axisymmetrical infiltration in soil imaged by non-invasive electrical resistivimetry, Soil Sci. Soc. Am. J., 73, 2, 510–520, 2009.

Birchak, J. R., C. G. Gardner, J. E. Hipp, and J. M. Victor, *High dielectric constant microwave probes for sensing soil moisture*, inProceedings of the IEEE, vol. 62, pp. 93–98, 1974.

Boivin, P., J. Touma, and P. Zante, Mesure de l'infiltrabilité du sol par la méthode du double anneau. 1- Résultats expérimentaux. *Cahiers Orstom, série Pédologique,* 24(1):17-25, 1987.

Clothier, B.E., and I. White, Measurement of sorptivity and soil water diffusivity in the field. S*oil Sci. Soc. Am. J.,* 45:241–245,1981.

Dane, J. H., and J. W. Hopmans, Hanging water column. *Soil Science Society of America*, Inc., Madison, WI., pp. 680–684, 2002.

Genuchten, M.T.V., A closed form equation for predicting the hydraulic conductivity of unsaturated soils, *Soil Sci. Soc. Am. J.*, vol. 44, pp. 892– 898, 1980.

Giannopoulos, A., Modelling ground penetrating radar by GprMax, *Construction and building materials*, vol. 19, no. 10, pp. 755–762, 2005.

Huisman, J.A., S. S. Hubbard, J. D. Redman, and A. P. Annan, Measuring soil water content with ground-penetrating radar: a review, *Vadose zone journal*, vol. 2, pp. 476–491, 2003.

Léger, E., and A. Saintenoy, Surface ground-penetrating radar monitoring of water infiltration inside a ring infiltrometer, In *14th International Conference on Ground Penetrating Radar*, Shanghai, 2012.

Müntz, A. ,L. Faure, and Laine E., Etudes sur la perméabilité des terres, faites en vue de l'arrosage, *Ann. De la Direction de l'Hydraulique*, vol. f33, pp. 45–53, 1905.

Perroux, K.M., and I. White, Designs for disc permeameters. *Soil Sci. Soc.Am. J.,* 52:1205–1215, 1988.

Reynolds, W.D., and D.E. Elrick, Determination of hydraulic conductivity using a tension infiltrometer. Soil Sci Sco. Am. J., 55, 633-639, 1991.

Saintenoy, A. and J. W. Hopmans, Ground Penetrating Radar: Water Table Detection Sensitivity to Soil Water Retention Properties, *Selected Topics in Applied Earth Observations and Remote Sensing, IEEE Journal of* , 4(4), 748,753, 2011

Saintenoy, A., S. Schneider, and P. Tucholka, Evaluating ground penetrating radar use for water infiltration monitoring, *Vadose zone journal*, vol. 7, pp. 208–214, 2008.

Šimůnek, J., and M.Th. van Genuchten, E*stimating unsaturated soil hydraulic properties from tension disc infiltrometer data by numerical inversion*. Water Resour. Res. 9:2683–2696, 1996.

Šimůnek, J., M. Šejna, and M.Th. van Genuchten. HYDRUS-2D. *Simulating water flow and solute transport in two-dimensional variably saturated media*. Version 1.0. Int. Ground Water Model. Ctr., Colorado School of Mines, Golden,1996.

Šimůnek, J., M. Th. van Genuchten, and M. Šejna, Development and applications of the HYDRUS and STANMOD software packages, and related codes, *Vadose Zone Journal*, *7*(2), 587-600, 2008.